\documentclass{iau}

\usepackage{amsmath}
\usepackage{graphicx}
\usepackage{multirow}

\begin{document}

\lefttitle{Shejeelammal et al.}
\righttitle{The [Y/Mg] chemical clock}

\jnlPage{1}{7}
\jnlDoiYr{2025}
\doival{10.1017/xxxxx}
\volno{395}
\pubYr{2025}
\journaltitle{Stellar populations in the Milky Way and beyond}

\aopheadtitle{Proceedings of the IAU Symposium}
\editors{J. Mel\'endez,  C. Chiappini, R. Schiavon \& M. Trevisan, eds.}

\title{The [Y/Mg] chemical clock and the stellar ages of solar-type stars in the solar neighbourhood}

\author{Shejeelammal Jameela$^{1}$, Jorge Mel{\'e}ndez$^{1}$, Anne Rathsam$^{1}$, Giulia Martos$^{1}$}
\affiliation{$^{1}$Universidade de São Paulo, Instituto de Astronomia, Geofísica e Ciências Atmosféricas, IAG, Departamento de Astronomia, \\ 
Rua do Matão 1226, Cidade Universitária, 05508-090, SP, Brazil}

\begin{abstract}
Stellar ages are an important parameter for studying the chemical evolution of the Galaxy. To improve age estimates, various complementary methods to the conventional isochrone-fitting method have been implemented in the last decade. The [Y/Mg] ratio has been proposed as a chemical clock for solar-metallicity stars, due to its strong dependence on stellar age. In this work, we study the behaviour of the abundance ratio [Y/Mg] in metal-poor stars, along with other solar-type disk stars, as a function of stellar age determined through isochrones. The analysis is based on the line-by-line differential technique, applied to a sample covering a wide metallicity range ($-$0.70 $\leq$ [Fe/H] $<$ +0.34), using high-resolution (R = 115,000) and high S/N spectra (= 270 - 1000) taken with the HARPS spectrograph at the ESO La Silla Observatory.
We find that the stellar sample follows a metallicity-dependent [Y/Mg]-Age relation.
\end{abstract}


\maketitle

\section{Introduction}
In addition to precise chemical abundances, the stellar age is an important parameter in studies of Galactic archaeology.
To better estimate the ages,various methods complementary to the conventional isochrone 
fitting method have been implemented in the past decade (e.g. \citealt{soderblom_2010}). One such method is the use of chemical clocks- abundance ratios 
that show a linear relation with stellar age (e.g. \citealt{dasilva_2012}). Generally, a pair of elements 
with different progenitors and opposite temporal evolution are chosen.
Over the past few years, the ratio between s-process and $\alpha$-elements ([s/$\alpha$]) has been found to be linearly 
related with age in Milky Way solar-type stars and giants, as well as in other galaxies (e.g. \citealt{nissen_2015, slumstrup_2017, skuladottir_2019, tautvai_2021, casali_2025}).
The [Y/Mg] clock is a promising technique, and it has been identified as a potential indicator of age in the case of solar twins (e.g. \citealt{nissen_2015, tuccimaia_2016, spina_2016}).
However, there are still several open questions, such as its validity for metal-poor stars and differences between the
thin and thick disk populations. Present work aims to study the metallicity dependence of the [Y/Mg] chemical clock using a 
sample of 233 solar-type disk stars with metallicity $-$0.70 $\leq$ [Fe/H] $<$ +0.34.

\section{Stellar sample and spectroscopic methodology}
Detailed procedure of sample selection, data processing, spectroscopic analysis techniques can be found in \cite{shejeelammal_2024}.
Here, we will briefly describe the relevant points. The stellar sample consisted of 233 solar type stars, and their 
high-resolution (R=115,000), high-quality (SNR= 270 -1000) HARPS spectra were obtained from the ESO archive. 
The stellar atmospheric parameters as well as the elemental abundances of Mg and Y were determined through a line-by-line differential 
analysis (e.g. \citealt{melendez_2012}) with respect to the Sun 
using the Python code \texttt{q$^{2}$} (qoyllur-quipu; \url{https://github.com/astroChasqui/q2}; \citealt{ramirez_2014}), 
adopting the line list from \cite{melendez_2014}. Stellar masses and ages of the sample were also determined from the probability distribution
functions through isochrone fitting. 

\section{Results and discussion}
\begin{figure}[t]
  \centerline{\vbox to 6pc{\hbox to 10pc{}}}
    \includegraphics[scale=0.4]{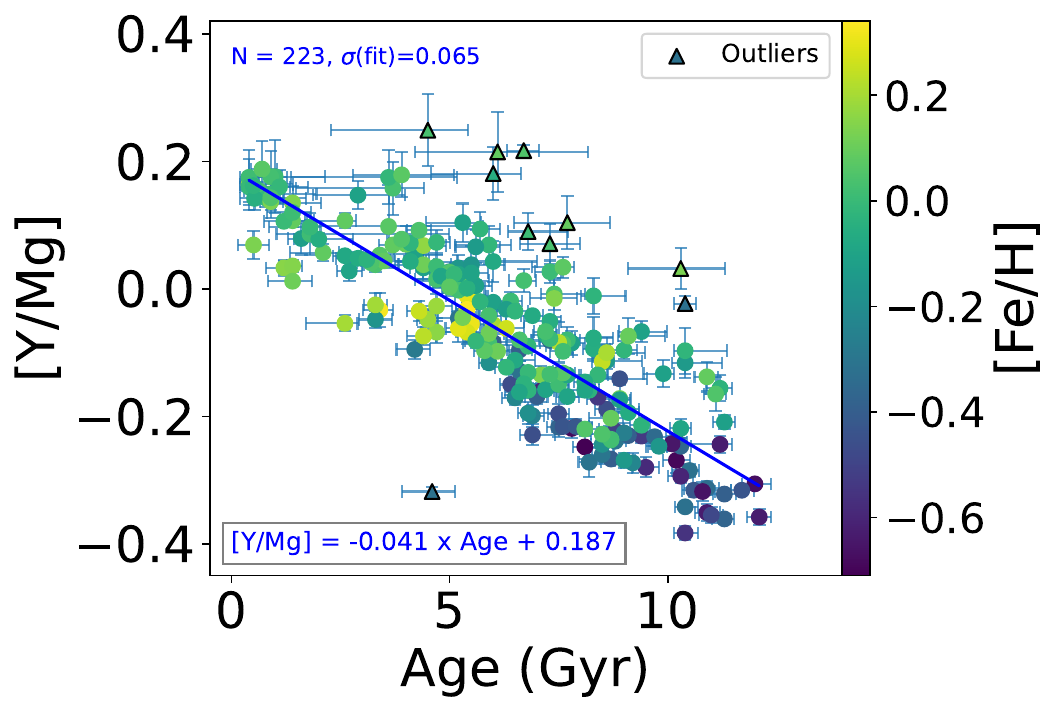}
     \includegraphics[scale=0.4]{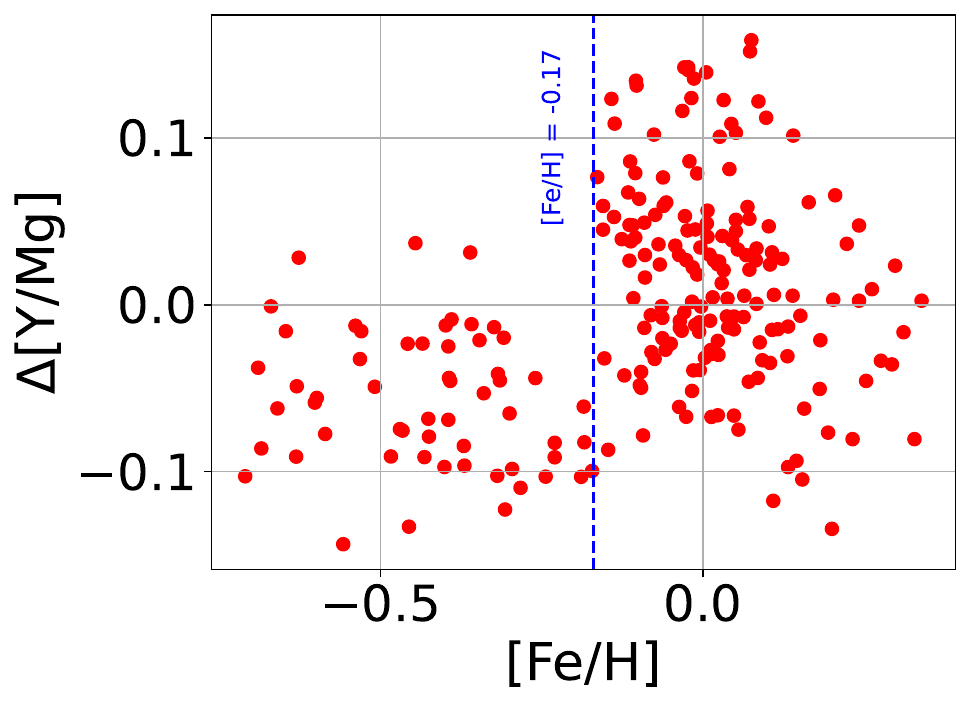}
  \caption{(i) Left panel: Best linear fit for the [Y/Mg]--Age distribution of the sample. (ii) right panel: distribution of the [Y/Mg] residuals ($\rm\Delta[Y/Mg]$) with respect to the metallicity of the sample. The vertical dashed line at [Fe/H] = $-$0.17 represents the point below which $\rm\Delta[Y/Mg]$ shows an offset with metallicity.} \label{YMg_age_residual}
\end{figure}

We found a strong [Y/Mg]--Age anti-correlation in our sample. To derive an empirical relation connecting [Y/Mg] and age, we performed a 
simple linear regression fitting of the form [Y/Mg] = a $\times$ Age + b. 
To get the best possible (accurate) relation, stars that lie outside the bulk of the sample were excluded (2$\sigma$ out of the trend). 
The relation obtained is shown in Fig. \ref{YMg_age_residual} (left panel).
To verify the metallicity dependence of this relation, the distribution of 
[Y/Mg] residuals {\footnotesize ($\rm\Delta[Y/Mg]$ = [Y/Mg]$_{obs}$ - [Y/Mg]$_{pred}$)} with respect to 
metallicity is evaluated (Fig. \ref{YMg_age_residual}, right panel). 
The $\rm\Delta[Y/Mg]$ is independent of metallicity up to a value of [Fe/H] $\sim$ $-$0.17 and show a trend below that, revealing 
the metallicity dependence of the [Y/Mg] chemical clock. 

Since we identified the metallicity dependence of the [Y/Mg] ratio in the sample, we also performed a multivariate (MV) linear regression fitting by 
incorporating [Fe/H] as well. The obtained relation as a function of stellar age and metallicity is given by: 
\begin{equation}
\rm
    [Y/Mg] = -0.036(\pm0.001) \times Age (Gyr) + 0.125(\pm0.010) \times [Fe/H] + 0.165(\pm0.034) \label{eq clock}
  \end{equation}
\noindent Eq. \ref{eq clock} is inverted to get a relation for stellar age (dating relation; Eq. \ref{dating relation}). 
 \begin{equation}
   \rm
   Age (Gyr) = -27.701(\pm0.461) \times [Y/Mg] + 3.454(\pm0.329) \times [Fe/H] + 4.576(\pm1.013) 
  \label{dating relation}
 \end{equation} 

\begin{figure}[t]
\centering
    \includegraphics[scale=0.4]{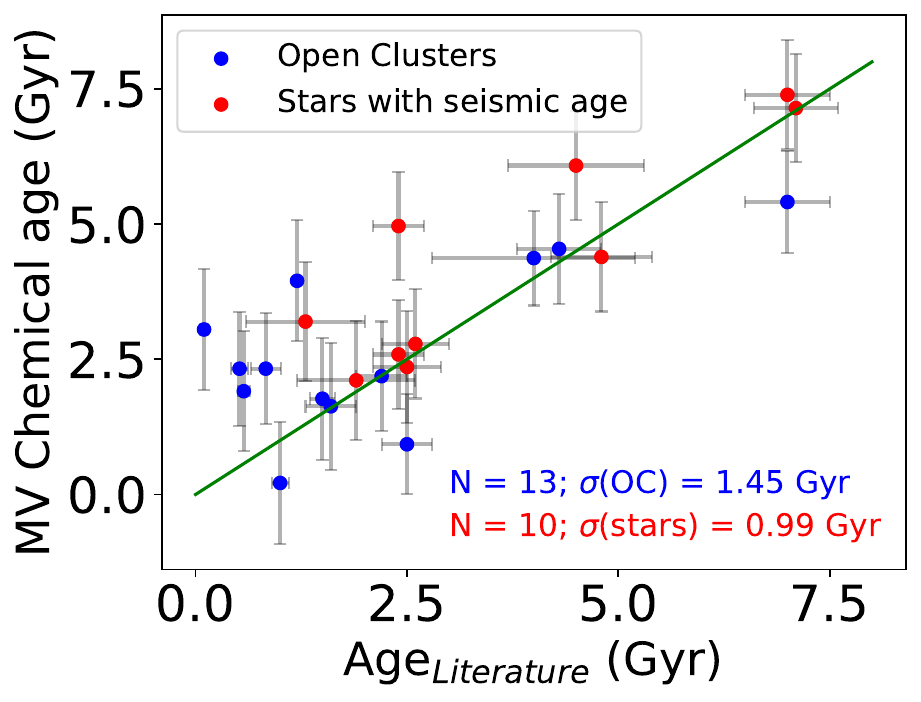}
  \caption{Comparison of the literature ages of the validation sample with their chemical ages calculated using Eq. \ref{dating relation}.} \label{Ocs_seismic_age_validation}
\end{figure}

We have validated our derived stellar dating relation using the data for open clusters (OCs) and 
stars with asteroseismic (KLS) ages (Fig. \ref{Ocs_seismic_age_validation}).
The accuracy we have obtained using our chemical clocks based on the KLS stars with 
asteroseismic ages and the OCs with isochrone ages respectively are $\sim$0.45 and 0.88 Gyr. 
These values are similar to the accuracy achieved for the 
solar twin stars in the solar neighbourhood ($\sim$0.8 - 1 Gyr; \citealt{nissen_2015, tuccimaia_2016}). Since the KLS stars 
are dwarfs and lie in the close vicinity of the Sun as our stellar sample, the validation using them is likely more accurate 
compared to the OCs.

\section{Conclusions}
The [Y/Mg] ratio shows a strong anti-correlation with stellar age. Our analysis confirms the metallicity dependence of the [Y/Mg] chemical clock. 
Using the dating relation (Eq. \ref{dating relation}), we achieve high accuracy and precision (0.45 and 0.99 Gyr, respectively). 
These values are compatible with the best accuracy achieved for solar twins to date ($\sim$0.8 - 1.0 Gyr) in the solar neighbourhood.
From our analysis, we find that the [Y/Mg]
chemical clock can be used as an age proxy for solar-type stars.
The high-precision outcomes of our
work are expected to provide contributions to the understanding
of different populations of the Galaxy and its chemical evolution.

\end{document}